\documentclass[aps,onecolumn,groupedaddress,nofootinbib,amsmath,amssymb]{revtex4}
\usepackage{dsfont}
\usepackage{bm}
\begin{document}
\setcounter{page}{1}
\title[]{Conformal symmetry of an extended Schr\"odinger equation and its relativistic origin}
\author{Mokhtar Hassa\"{\i}ne}\email{hassaine-at-inst-mat.utalca.cl}
\affiliation{Instituto de Matem\'atica y F\'{\i}sica, Universidad de
Talca, Casilla 747, Talca, Chile,} \affiliation{Centro de Estudios
Cient\'{\i}ficos (CECS),
 Casilla 1469, Valdivia, Chile.}

\begin{abstract}
In this paper two things are done. We first prove that an arbitrary
power $p$ of the Schr\"odinger Lagrangian in arbitrary dimension
always enjoys the non-relativistic conformal symmetry. The
implementation of this symmetry on the dynamical field involves a
phase term as well as a conformal factor that depends on the
dimension and on the exponent. This non-relativistic conformal
symmetry is shown to have its origin on the conformal isometry of
the power $p$ of the Klein-Gordon Lagrangian in one higher dimension
which is related to the phase of the complex scalar field.

\end{abstract}

\maketitle

\section{Introduction}

When a law of physics does not change against some transformations,
the system is said to exhibit some symmetries. The symmetries of a
physical system are given by the transformations that do not change
the mathematical structure of the system. One can even start by
defining the transformations and then to find the mathematical
structure compatible with these transformations.  The determination
of the symmetries of a system can also be a powerful instrument
since it may allow to put the problem into a simpler form or it can
permit to obtain nontrivial solutions from trivial ones. It is then
clear that the problem of the identification of the symmetries
underlying an equation is not an academic question but rather a
fundamental one. In this paper we shall be concerned with the
Schr\"odinger symmetry that is defined as the dynamical symmetry
leaving invariant the free Schr\"odinger equation, see
\cite{Schrodinger}, \cite{Schrodinger1} and \cite{Schrodinger2}. The
Schr\"odinger invariance has been relevant in a wide variety of
situations as celestial mechanics \cite{Duval:1990hj},
non-relativistic field theory \cite{J}-\cite{Jackiw:1990mb},
non-relativistic quantum mechanics \cite{Ghosh:2001an},
hydrodynamics
\cite{Jackiw:2000mm,Hassaine:1999hn,O'Raifeartaigh:2000mp,Hassaine:2000ti},
in the context of the AdS/CFT correspondence \cite{Leiva:2003kd} as
well as in statistical physics \cite{Henkel:2006wm}.

The Schr\"odinger group is defined  as the largest group of
space-time transformations which leave invariant the free
Schr\"odinger equation
\begin{equation}
\left[i\partial_t+\frac{1}{2}\Delta_d\right]\Phi=0, \label{schrod}
\end{equation}
where the operator $\Delta_d$ representes the Laplacian in $d$
spatial dimensions. In $d+1$ dimensions, the Schr\"odinger group is
a $[d(d+3)+6]/2-$dimensional Lie group which can be viewed as the
semi direct sum of the static Galilei group with the $SL(2,I\!\! R)$
group. The static Galilei group which is a $d(d+3)/2-$parameter
group induces the static Galilei transformations given by
\begin{eqnarray}
t\to t,\qquad \qquad \vec{x}\to {\cal R}\vec{x}+\vec{\chi}-\vec{v}t,
\label{galileis}
\end{eqnarray}
where ${\cal R}\in SO(d)$, $\vec{\chi}\in \mathds{R}^d$ and
$\vec{v}\in \mathds{R}^d$ generate respectively the rotations, the
spacial translations and the Galilean boosts. On the other hand,
$SL(2,I\!\! R)$ is the group which induces the following
transformations
\begin{eqnarray}
t\to \tilde{t}=\frac{\alpha t+\beta}{\gamma
t+\delta},\qquad\qquad\vec{x}\to
\vec{\tilde{x}}=\frac{\vec{x}}{\gamma t+\delta}, \label{sl2r}
\end{eqnarray}
and is a three-parameter group since the parameters are tied by the
relation $\alpha\delta-\beta\gamma=1$. These transformations include
the time translations ($\gamma=0$, $\alpha=\delta=1$), the
dilatations ($\beta=\gamma=0$) and the special conformal
transformations also called the expansions ($\alpha=\delta=1$,
$\beta=0$). The static Galilei transformations and time translations
induce the following field change
\begin{eqnarray}
\tilde{\Phi}(t,\vec{x})=e^{i
(\vec{x}\cdot\vec{v}-\frac{\vert\vec{v}\vert^2
t}{2})}\,\Phi(t+\beta,{\cal R}\vec{x}+\vec{\chi}-\vec{v}t),
\label{gc}
\end{eqnarray}
while the scalar field changes under the dilatation as
\begin{eqnarray}
\tilde{\Phi}(t,\vec{x})=\alpha^{d/2}\Phi(\alpha^2t, \alpha\vec{x}),
\label{gcc}
\end{eqnarray}
and under the special conformal transformations as
\begin{eqnarray}
\tilde{\Phi}(t,\vec{x})=\frac{1}{(1+\gamma t)^{d/2}}\,\,e^{\frac{i
\gamma\vert\vec{x}\vert^2}{2(1+\gamma t)}}\,\,\Phi(\frac{t}{1+\gamma
t}, \frac{\vec{x}}{1+\gamma t}). \label{gccc}
\end{eqnarray}
We remark that in the case of the the dilatation and the expansion,
the transformation of the dynamical field is associated with a
multiplicative factor given by $J^{\frac{d}{2(d+2)}}$ where $J$ is
the Jacobian of the transformation linking
$(\tilde{t},\vec{\tilde{x}})\to (t,\vec{x})$. In order to be
complete, we also recall that the free Schr\"{o}dinger equation
(\ref{schrod}) is derived from the following Lagrangian
\begin{eqnarray}
{\cal
L}_S=-\frac{i}{2}\left(\Phi^{\star}\partial_t\Phi-\Phi\partial_t\Phi^{\star}\right)+
\frac{1}{2}\vert\vec{\nabla}\Phi\vert^2 \label{la}
\end{eqnarray}
which enjoys the Schr\"{o}dinger symmetry as well.

The plan of the paper is organized as follows. In the next part, we
show that an arbitrary power $p$ of the Schr\"{o}dinger Lagrangian
(\ref{la}) in $(d+1)$ dimensions also enjoys the non-relativistic
conformal symmetry. The associated Noether conserved quantities are
derived and are shown to reduce to the standard Schr\"{o}dinger
quantities for $p=1$. The origin of this non-relativistic conformal
symmetry is explained in the Section $3$ using a Kaluza-Klein type
framework in one higher dimension; this extra dimension is related
with the phase of the complex scalar field. More precisely, we
define a $(d+2)-$dimensional Minkowski spacetime endowed with a
covariantly constant and lightlike vector field $\xi$. On this
manifold, we consider the relativistic action given by the power $p$
of the complex Klein-Gordon Lagrangian, and we show that the
conformal isometries preserving the vector $\xi$ are precisely the
non-relativistic symmetries of the extended Schr\"{o}dinger
equation.

\section{Non-relativistic conformal equation}
We now consider an action in $(d+1)$ dimensions defined as an
arbitrary power $p$ of the Schr\"{o}dinger Lagrangian (\ref{la})
\begin{eqnarray}
S_p=\int d^d\vec{x}\,dt\,{\cal L}_S^p=\int
d^d\vec{x}\,dt\,\left(-\frac{i}{2}\left(\Phi^{\star}\partial_t\Phi-\Phi\partial_t\Phi^{\star}\right)+
\frac{1}{2}\vert\vec{\nabla}\Phi\vert^2\right)^{p}, \label{action}
\end{eqnarray}
where $p$ is a real parameter. The associated field equation reads
\begin{eqnarray}
\label{dec} p\Big(i\partial_t\Phi+\frac{1}{2}\Delta_d\Phi\Big){\cal
L}_S^{p-1} +\frac{1}{2}p(p-1)\left(i\Phi\,(\partial_t{\cal
L}_S)+\vec{\nabla}\Phi\cdot\vec{\nabla}{\cal L}_S\right){\cal
L}_S^{p-2}=0,
\end{eqnarray}
and reduces to the standard Schr\"{o}dinger equation for $p=1$. For
later convenience, we derive from (\ref{dec}) the continuity like
equation given by
\begin{eqnarray}
\partial_t\left(\vert\Phi\vert^2{\cal
L}_S^{p-1}\right)+\vec{\nabla}\cdot\left(\frac{-i}{2m}
(\Phi^{\star}\vec{\nabla}\Phi-\Phi\vec{\nabla}\Phi^{\star}){\cal
L}_S^{p-1}\right)=0. \label{ce}
\end{eqnarray}

We now show that for any arbitrary value of $p\not=0$, the extended
equation (\ref{dec}) or equivalently the action (\ref{action})
possess the Schr\"{o}dinger symmetry. Firstly, the space-time
transformations leaving invariant the extended equation are given by
the usual Schr\"{o}dinger transformations (\ref{galileis}) and
(\ref{sl2r}). The main difference lies in the implementation of the
conformal transformations on the dynamical field $\Phi$. Indeed, the
action of the static Galilei transformations and time translations
is the same as in the standard Schr\"{o}dinger case (\ref{gc}), but
the scalar field changes under the dilatation ($\gamma=0$) and the
special conformal transformation ($\alpha=1$) as
\begin{eqnarray}
\tilde{\Phi}(t,\vec{x})=\left(\frac{\alpha}{1+\gamma
t}\right)^{\frac{d+2-2p}{2p}}\,e^{i\frac{\gamma\vert\vec{x}\vert^2}{2(1+\gamma
t)}}\,\,\Phi(\frac{\alpha^2\, t}{\gamma t+1},\frac{\alpha\,
\vec{x}}{\gamma t+1}) \label{ngcc}
\end{eqnarray}
This means that for a scalar field $\Phi(t,\vec{x})$ solving the
field equation (\ref{dec}), then so also do the transformed fields
$\tilde{\Phi}(t,\vec{x})$ defined by (\ref{ngcc}) for any value of
the parameter $p$. The same conclusion can be obtained by observing
that under a dilatation or a special conformal transformation
(\ref{ngcc}), the Schr\"{o}dinger Lagrangian rescales as
$$
{\cal L}_S\to \left(\frac{\alpha}{1+\gamma t}\right)^{(d+2)/p}{\cal
L}_S,
$$
and hence the power $p$ of the Schr\"{o}dinger Lagrangian exactly
compensates the Jacobian of the conformal transformations which
means that the action (\ref{action}) remains unchanged  $S_p\to
S_p$. A direct application of the Noether theorem yields the
following constants of motion
\begin{eqnarray}
\label{qcon} \label{energy} &H&=\int d^{d}\vec{x}\,\,{\cal H}=\int
 d^{d}\vec{x}\,\left[\frac{p}{2}\vert\vec{\nabla}\Phi\vert^2\,{\cal
 L}_S^{p-1}\right],\nonumber\\
\label{momentum} &\vec{P}&=\int d^{d}\vec{x}\,\,\vec{{\cal P}}=\int
d^{d}\vec{x}\, \left[-\frac{ip}{2}
\left({\Phi}^{\star}\vec{\nabla}\Phi-\Phi\vec{\nabla}\Phi^{\star}\right){\cal
L}_S^{p-1}\right],\nonumber\\ \label{rotation} &M_{ij}&=\int
d^{d}\vec{x}\,\,\left(x_i{\cal
P}_j-x_j{\cal P}_i\right),\\
\label{boosts} &\vec{G}&=t\vec{P}-p\,\int
d^{d}\vec{x}\,\,\vert\Phi\vert^2{\cal L}_S^{p-1},\nonumber\\
&D&=t\,H-\frac{1}{2}\int d^d\vec{x}\,\,\left(\vec{x}\cdot\vec{{\cal
P}}\right),\nonumber\\
&K&=-t^2H+2tD+\frac{p}{2}\int
d^d{\vec{x}}\,\left(\vert\vec{x}\vert^2\vert\Phi\vert^2{\cal
L}_S^{p-1}\right)\nonumber,
\end{eqnarray}
which correspond respectively to the energy (time translation), the
momentum (space translations), the rotations, the Galilean boosts,
the dilatation and the special conformal transformation. Note that
the conservation of the functional $K$ can be viewed as a
consequence of the conservations of the energy and dilatation
functionals together with the continuity equation (\ref{ce}). We
also mention that more general non-linear terms yielding
Schr\"{o}dinger invariant equations can also be considered, see
\cite{Stoimenov:2005aa} and \cite{Baumann:2005gk}.

Various comments can be made at this stage of the analysis. Firstly,
for $p=1$, all the expressions derived reduce to those associated to
the standard Schr\"{o}dinger theory. Secondly, the invariance of the
extended action is achieved for any value of the power $p$. This is
due to the fact that the Jacobian of the conformal transformation
can always be compensated by rescaling in an appropriate way the
dynamical field (\ref{ngcc}). For the particular exponent
$p=(d+2)/2$, the multiplicative factor different from the phase term
in the field transformation (\ref{ngcc}) is not longer present. This
means that the Jacobian of the conformal transformation is exactly
compensated by the power $p=(d+2)/2$ of the free Schr\"{o}dinger
Lagrangian without necessity of rescaling the dynamical field but
just by operating a phase change. In the next section, we will
explain the origin of this extended non-relativistic Schr\"{o}dinger
symmetry within a higher-dimensional relativistic context.

\section{Relativistic origin}
The purpose of this section is to provide an explanation of the
Schr\"{o}dinger symmetry of the extended action using a relativistic
framework in one higher dimension. The clue of this relativistic
framework lies in the fact that non-relativistic space-time $Q$ in
$(d+1)$ dimensions can be viewed as the quotient of a
$(d+2)-$dimensional Lorentz manifold $M$ by the integral curves of a
covariantly constant, lightlike vector field $\xi$. This
correspondence has been used in order to derive the Schr\"{o}dinger
symmetry of the standard Schr\"{o}dinger equation from the
relativistic conformal symmetry of the conformal wave equation, see
\cite{Duval:1984cj}, \cite{Duval:1990hj} and \cite{Duval:1994pw}.

On the manifold $M$ we adopt the coordinate system $(t,\vec{x}_d,s)$
where $(t,\vec{x}_d)$ are the coordinates on $Q$ and $s$ is the
additional coordinate, and we consider the $(d+2)-$dimensional
action given by the power $p$ of the Klein-Gordon Lagrangian
\begin{eqnarray}
{\cal S}_p=\int_M \sqrt{-g}\,d^{d+2}x\,\left[\frac{1}{2}g^{\mu\nu}
\partial_{\mu}\psi\partial_{\nu}\psi^{\star}\right]^p.
\label{relacza}
\end{eqnarray}
The field equation obtained by varying this action with respect to
the complex scalar field yields
\begin{eqnarray}
\frac{1}{\sqrt{-g}}\partial_{\mu}\left[\sqrt{-g}\,\partial^{\mu}\psi
\left(\partial_{\sigma}\psi\partial^{\sigma}\psi^{\star}\right)^{p-1}\right]=0.
\label{bnm}
\end{eqnarray}
On the Minkowski spacetime, we consider the flat metric written in
lightcone coordinates as
\begin{eqnarray}
ds^2=d\vec{x_d}^2+2dtds \label{lcm}
\end{eqnarray}
for which the covariantly constant, lightlike vector field $\xi$ is
chosen to be $\xi^{\mu}\partial_{\mu}=\partial_s$. In order to
establish the correspondence with the extended Schr\"{o}dinger
equation (\ref{dec}), the field $\psi$ is assumed to satisfy an
equivariance condition given by
\begin{eqnarray}
\xi^\mu\partial_\mu\psi=i\psi, \label{equi}
\end{eqnarray}
which in turn implies that the function
\begin{eqnarray}
\Phi=e^{-is}\psi \label{serd}
\end{eqnarray}
is a function defined on $Q$ since $\partial_s\Phi=0$. It is then
simple to see that in Minkowski space with the metric (\ref{lcm}),
the extended wave equation (\ref{bnm}) together with the
equivariance condition (\ref{equi}) are equivalent to the extended
Schr\"odinger equation (\ref{dec}). The same conclusion can be
achieved at the level of the actions in the sense that the action
(\ref{relacza}) evaluated on the flat metric (\ref{lcm}) for a
scalar field satisfying the equivariance condition (\ref{equi})
reduces to the extended Schr\"odinger action (\ref{action}). We now
study the symmetries of the coupled system given by the equation
(\ref{bnm}) with the condition (\ref{equi}). We recall that a
conformal isometry on the Lorentz manifold $M$ is a diffeomorphism
$\varphi:M\to M$ for which there is a nonvanishing function $\Omega$
such that $(\varphi^{\star}g)_{\mu\nu}=\Omega^2\,g_{\mu\nu}$, see
e.g. \cite{Wald}. It is simple to show that the conformal isometries
that preserve the vertical vector, i.e. $\varphi^{\star}\xi=\xi$,
are symmetries of the coupled system (\ref{bnm}) and (\ref{equi})
for which the implementation on the dynamical field is given by
\begin{eqnarray}
\label{cf2}
\tilde{\psi}=\Omega^{\frac{2p-d-2}{2p}}\,\,\varphi^{\star}\psi.
\end{eqnarray}
In the case of the flat metric (\ref{lcm}), the $\xi$-preserving
conformal isometries transformations form a subgroup of the
conformal group and the conformal Killing vector field is given in
the basis $(\partial_t,\vec{\partial},\partial_s)$ by
\begin{eqnarray}
(X^\mu)= \left(
\begin{array}{c}
\chi t^2+\delta t+\epsilon\\
\\
{\cal R}\vec{x}+(\frac{1}{2}\delta+\chi t)\vec{x}+t\vec{\beta}+\vec{\gamma}\\
\\
-\frac{1}{2}\chi \vert\vec{x}\vert^2-\vec{\beta}\cdot\vec{x}+\eta
\end{array}
\right), \label{lengard}
\end{eqnarray}
where ${\cal R}\in SO(d)$, $\vec{\beta}, \vec{\gamma}, \epsilon,
\chi, \delta$ and $\eta$ are interpreted as rotation, boost, space
translation, time translation, expansion, dilatation and vertical
translation. The integration of the Lie differential equation for
transformation group yields to the following spacetime
transformations
\begin{subequations}\label{dfgs}
\begin{eqnarray}
\label{qa1}
&&\tilde{\vec{x}}=\frac{{\cal R}\vec{x}-\vec{v}t+\vec{\chi}}{\gamma t+\delta},\\
 \label{qa2}
&&\tilde{t}=\frac{\alpha t+\beta}{\gamma t+\delta},\\
\label{qa3} && \tilde{s}=s+\frac{\gamma}{2}\frac{\left\vert{\cal
R}\vec{x}-\vec{v}t+\vec{\chi}\right\vert^2}{\gamma t+\delta}+{\cal
R}\vec{x}\cdot\vec{v}-\frac{t}{2}\vert\vec{v}\vert^2+h,
\end{eqnarray}
\end{subequations}
with the restriction $(\alpha\delta-\beta\gamma)=1$. The
corresponding conformal isometry factor is given by
\begin{eqnarray}
\Omega=\Omega(t)=\gamma t+\delta.\label{al}
\end{eqnarray}
The first two equations (\ref{qa1}-\ref{qa2}) correspond to the
Schr\"odinger transformations while the third transformation
(\ref{qa3}) possesses the information concerning the phase change of
the complex scalar field. Indeed, combining the equivariance
condition (\ref{equi}-\ref{serd}) together with the law
transformation of $\psi$ (\ref{cf2}), we have
\begin{eqnarray}
\tilde{\psi}(t,\vec{x},s)=e^{is}\,\tilde{\Phi}(t,\vec{x})
\Longrightarrow
\tilde{\Phi}(t,x)=\Omega(t)^{\frac{2p-d-2}{2p}}\,\,e^{i(\tilde{s}-s)}\,
 \Phi(\tilde{t},\tilde{\vec{x}})
\end{eqnarray}
and we obtain the change field for the extended Schr\"odinger field
(\ref{ngcc}). More precisely, from this expression, it is clear that
the phase change of the Schr\"{o}dinger field is associated with the
change of the additional coordinate $s$ while the multiplicative
factor is given by the conformal isometry factor (\ref{al}).

To conclude we mention that the same analysis can be done in curved
spacetime by considering the following action
\begin{eqnarray}
L_{p}=\int
\sqrt{-g}\,d^{d+2}x\,\Big[\frac{1}{4}(2p-d-2)\left(\psi\Box\psi^{\star}+\psi^{\star}\Box\psi\right)
+\frac{(2p-d-2)^2}{8p(d+1)}R\vert\psi\vert^2
-\frac{1}{2}(d+2)(p-1)g^{\mu\nu}\partial_{\mu}\psi
\partial_{\mu}\psi^{\star}\Big]^p,
\label{relacp}
\end{eqnarray}
which enjoys, for any value of the parameter $p$, the conformal
invariance with weight given by
\begin{eqnarray}
g_{\mu\nu}\to \Omega^2g_{\mu\nu},\qquad \qquad \psi\to
\Omega^{\frac{2p-d-2}{2p}}\psi. \label{pesop}
\end{eqnarray}
This action generalizes the standard conformal wave action since for
$p=1$, the conformal extended action (\ref{relacp}) reduces after an
integration by parts to the conformal wave action. It is also
interesting to note that for the particular exponent $p=(d+2)/2$,
the conformal action reduce to the standard kinetic term to this
power. In this case, the conformal symmetry can be viewed as the
higher-dimensional extension of the two-dimensional conformal
Klein-Gordon action \cite{Hassaine:2005xg}. Note that the
two-dimensional situation is very special since in this case, the
conformal algebra is the direct sum of two isomorphic
infinite-dimensional algebras \cite{DiFrancesco:1997nk}.

Finally, it is also legitimate to wonder about the non-relativistic
limit which is a quite difficult question \cite{DeMontigny:2005rm}.
In the standard case $p=1$, it has been shown by direct computation
that the resulting Lie algebra is not the Schr\"{o}dinger algebra
but a different algebra of same dimension and not isomorphic to the
Schr\"{o}dinger algebra \cite{Henkel:2003pu}. For $p\not=1$, the
same conclusion will still be valid since the algebras involved are
the same.

\section{Summary and discussion}
Here, we have shown that the non-relativistic conformal symmetry of
the Schr\"{o}dinger Lagrangian is still valid for any power of the
Schr\"{o}dinger Lagrangian. More precisely, the spacetime
transformations leaving invariant the extended action are the usual
Schr\"{o}dinger transformations but the main difference lies in the
implementation of the conformal transformations on the dynamical
field. Indeed, this implementation is realized through a conformal
factor that depends on the dimension and on the exponent as well as
with a phase term. There exists a particular value of the exponent
for which this conformal factor is not longer present. For a generic
value of the exponent, the associated Noether conserved quantities
have been obtained. The origin of this non-relativistic conformal
symmetry has been analyzed within a relativistic framework in one
higher dimension. The same conclusions may also be valid by
considering an arbitrary power of the Newton-Hooke Lagrangian
because of the various analogies between both models. The main
difference lies in the fact that the conformal symmetry of the free
Schr\"odinger equation is associated to the conformal isometries
that preserve the vertical vector in flat space while in the
Newton-Hooke context, the metric is an homogenous plane wave metric.
This is due to the fact that the Newton-Hooke group can be obtained
from the (A)dS groups as the non-relativistic limit with the
velocity of light $c$ going to infinity and the cosmological
constant $\Lambda$ going to zero while keeping $c^2\Lambda$ finite,
see e.g. \cite{Gibbons:2003rv} and for recent work
\cite{Alvarez:2007fw}.

The conformal invariance of the relativistic action (\ref{relacp})
independently of the power $p$ may be interesting in the search of
black hole solutions. Indeed, in the standard case $p=1$ and in four
dimensions, the Einstein equations with this conformal source admits
black hole solutions \cite{BBMB,BEK}. In this example, the conformal
character of the matter source has been crucial since the solution
has been derived using the machinery of conformal transformations
applied to minimally coupled scalar fields \cite{BEK}. It would be
interesting to explore whether there exist black hole solutions for
the Einstein equations with the conformal source (\ref{relacp}).

\bigskip

\acknowledgments  We thank Eloy Ay\'on-Beato and Mikhail Plyushchay
for useful discussions. This work is partially supported by grants
1051084 and 1060831 from FONDECYT. Institutional support to the
Centro de Estudios Cient\'{\i}ficos (CECS) from Empresas CMPC is
gratefully acknowledged. CECS is a Millennium Science Institute and
is funded in part by grants from Fundaci\'{o}n Andes and the Tinker
Foundation.


\end{document}